\documentclass[aip,
 amsmath,amssymb,
 reprint,%
]{revtex4-1}
\usepackage{amsmath}
\usepackage{amsfonts}
\usepackage{amssymb}
\usepackage{dcolumn} %% tables cols aligned at decimal point
\usepackage{array}
\usepackage{color}
\usepackage{float}
\usepackage{psfrag}
\usepackage{graphicx}
\usepackage{braket}
\usepackage{rotating}
\usepackage{subfig}
\usepackage{epsfig}
\usepackage{adjustbox}
\usepackage{ragged2e}
\usepackage{simplewick}
\usepackage{xcolor}
\usepackage{fancyhdr}
\usepackage{multirow}
\usepackage{csquotes}
\usepackage{gensymb}
\usepackage{url}

\usepackage{bm}% bold math
\usepackage[utf8]{inputenc}
\usepackage[T1]{fontenc}
\usepackage{mathptmx}

\begin{document}

\preprint{AIP/123-QED}

 \author{Valay Agarawal}
  \author{Anish Chakraborty}
 %\affiliation{IITB write full name and address}
  \author{Rahul Maitra}
  \email{rmaitra@chem.iitb.ac.in}
 \affiliation{Department of Chemistry, Indian Institute of Technology Bombay, Powai, Mumbai 400076, India} 
\title {Stability analysis of a double similarity transformed coupled cluster theory}

\date{\today}% It is always \today, today,

\begin{abstract}
In this paper, we have analysed the time series associated with the iterative scheme of a double similarity transformed Coupled Cluster theory. The coupled iterative scheme to solve the ground state Schr{\"o}dinger equation is cast as a multivariate time-discrete map, the solutions show the universal Feigenbaum dynamics. Using recurrence analysis, it is shown that the dynamics of the 
iterative process is dictated by a small subgroup of cluster operators, mostly those involving chemically 
active orbitals, whereas all other cluster operators with smaller amplitudes are enslaved. 
Using Synergetics, we will indicate how the master-slave dynamics can suitably 
be exploited to develop a novel coupled-cluster algorithm in a much reduced dimension.

\end{abstract}

\maketitle
\section{Introduction and Theory}
\label{section: Introduction}
%\textbf{\textit{Introduction and Theory:}} 
Coupled Cluster (CC) theory\cite{cc3,cc4,cc5,bartlett2007coupled} is an accurate electronic structure 
methodology to compute the energetics and properties of small to medium-sized atoms and molecules. CC 
theory, with singles, doubles, and perturbative triples excitations, the so-called 
CCSD(T)\cite{ccsdpt1,ccsdpt2,ccsdpt3} is known to provide very accurate results for molecules in their 
near-equilibrium geometry. Recently, a new iterative scheme, known as the \textit{iterative n-body
excitation inclusive CCSD, (iCCSDn)}\cite{maitra_coupled_2017,maitra_correlation_2017}, 
which takes care of the fully connected triple excitations at a computational cost less than 
that of CCSD(T) has been proposed. Owing to the nonlinear nature of the CC amplitude determining equations, one employs an
iterative procedure to find the solutions, which are the fixed points of the iterative process. 
Here we will present a posteriori analysis of the time series associated 
with the iterative process of iCCSDn methodology.
By introducing a regularization parameter to probe the non-linearity associated with the dynamics of
the iteration process, we will show that such an iteration process exhibits 
rich dynamical features. Furthermore, a detailed study of the dynamics
would help us to establish certain interrelationships among the different 
cluster operators. One may further exploit the synergy 
among different cluster operators to construct a novel CC iterative scheme where the effective degrees 
of freedom is drastically reduced. Thus the study of the dynamics associated with the CC iteration
process is important from a quantum chemistry perspective. Such a scheme to formulate an effective 
CC theory with significantly fewer iterables will sketchily be presented towards the end of the article. 
We note that such an analysis holds true for 
conventional CC theory as well; however, we will restrict our analysis solely to iCCSDn.

In iCCSDn, one parametrizes the wavefunction as a double exponential waveoperator $\Omega$ acting on a reference zeroth-order wavefunction, usually taken to be the Hartree-Fock (HF) determinant.
\begin{equation}
\Omega = \{\exp(S)\}\exp (T_1+T_2)
\label{eq2}
\end{equation}
where $T$'s are the usual CCSD excitation operators (also known as the cluster operators), and $S$ denote scattering operators that induce higher excitations by their action on the doubly excited determinants. $S$ and $T$ operators do not commute. The presence of hole$\to$hole (or particle$\to$particle) scattering in $S$ ensures that its action on HF determinant is trivially zero, but not on an excited determinant. The higher rank correlation effect is simulated via the contraction of $S$ and $T$ operators, and hence it provides the accuracy at a cheap computational scaling. The quantity inside $\{...\}$ denotes 'Normal Ordering,' which ensures the CC expansion terminates at finite power. The effective Hamiltonian, $G$, is constructed via two similarity transformations recursively. 
\begin{equation}
G = e^{-(T_1+T_2)} W e^{(T_1+T_2)}
\end{equation}
where,
    $W=\{\contraction{}{H}{}{\exp(S)} H \exp(S)  - {\contraction[2ex] {}{(\exp(S)-1)}{}{\contraction{}{H}{}{\exp(S)} H \exp(S)} (\exp(S)-1) \contraction{}{H}{}{\exp(S)} H \exp(S)} \}$
is the first similarity transformed Hamiltonian obtained through the time-independent Wick's theorem and the connections depict Wick contraction. The determination of the cluster operators can be done in a coupled manner at a scaling marginally higher than CCSD. The cluster operator $T$'s are responsible for inducing the dynamical correlation, whereas the $S$ operators renormalize them through a set of local denominators, by including the effects of connected triple excitations within the two-body cluster amplitudes. Thus, they are expected to be large at stretched molecular geometries. Note that the $S$ operators do not have any direct effect on energy; however they do indirectly contribute at high perturbative orders by renormalizing the cluster amplitudes.

Following a many-body expansion\cite{nooijen2000} of the double similarity transformed Hamiltonian, $G$, the amplitudes $t$($s$) associated with $T$($S$) and  operators are obtained through a set of coupled non-linear equations by demanding $g_\mu = g_\alpha=0$ upon convergence. Here $g$ is the amplitude associated with the tensor $G$ and $\mu$ ($\alpha$) are the collective orbital labels associated with the tensor $T$ ($S$). Let us denote the orbital labels associated with $T$ as $\mu, \nu,...$, etc. and those associated with the scattering operator $S$ as $\alpha, \beta,...$ etc. In the iteration procedure, the discrete-time propagation of vector at $(n+1)$-th step can be represented as 
time-discrete maps:
\begin{eqnarray}
t_{\mu}^{(n+1)} = t_{\mu}^{(n)} + \frac{g_{\mu}}{D_{\mu}+\eta} = f_\mu(t^{(n)},s^{(n)}) \nonumber \\
s_{\alpha}^{(n+1)} = s_{\alpha}^{(n)} + \frac{g_{\alpha}}{D_{\alpha} + \eta} = f_\alpha(t^{(n)},s^{(n)})
\label{eq1}
\end{eqnarray} 
Here $D$ is a suitably chosen denominator, usually taken as the HF orbital energy difference associates with the orbital labels of $\mu$ (and $\alpha$), and $\eta$ is known as the damping or 
regularization parameter, which is often used to accelerate the convergence of the iterative procedure
without affecting the fixed points. Note that in control theory and system engineering, such an 
external parameter is known as input, perturbation or control parameter. Following the standard
nomenclature in quantum chemistry, we will call $\eta$ as the regularization parameter which controls 
the dynamics of the iteration process. Let vector $(\tilde{t}, \tilde{s})$ be the fixed 
points of the equation,
such that $t_\mu = f_\mu(\tilde{t},\tilde{s})$ and $s_\alpha = f_\alpha(\tilde{t},\tilde{s})$, or in general $(t_\mu,s_\mu)= f(\tilde{t},\tilde{s})$. Here $f_\mu$ and
$f_\alpha$ are the functions having the same hole/particle tensor structure as $g_\mu$ and 
$g_\alpha$ respectively and $f$ is the generic symbol of $f_\mu$ and $f_\alpha$. 
Following Sur{\'j}an \cite{szakacs2008stability,szakacs2008iterative}, let's assume a small deviation around the fixed points to be $\xi$ such that
\begin{equation}
(t^{(n)},s^{(n)}) = (\tilde{t},\tilde{s}) + \xi^{(n)}
\end{equation}

So, employing Taylor series expansion around the fixed points, we obtain-
\begin{equation}
(t_\mu,s_\mu) + \xi^{(n+1)} = f(\tilde{t},\tilde{s}) + \sum_{\nu=1}^{m} \frac{\partial f}{\partial t_{\nu}}\bigg|_{\tilde{t},\tilde{s}}\xi_{\nu}^{(n)}
+ \sum_{\beta=1}^{m} \frac{\partial f}{\partial s_{\beta}}\bigg|_{\tilde{t},\tilde{s}}\xi_{\beta}^{(n)} + \cdot\cdot\cdot
\label{eq3}
\end{equation}

Note from Eq. (\ref{eq1}) that $f$ is a function of 
both $t_\nu$ and $s_\beta$ amplitudes. Assuming a small perturbation 
such that the terms $O({\xi^{(n)}}^2)$ are negligible, one may write Eq. (\ref{eq3})
in the long-hand notation as:
\begin{equation}
\xi_{\mu}^{(n+1)} = \sum_{\nu=1}^{m} \frac{\partial f_{\mu}}{\partial t_{\nu}}\bigg|_{\tilde{t},\tilde{s}}\xi_{\nu}^{(n)}
+ \sum_{\beta=1}^{m} \frac{\partial f_{\mu}}{\partial s_{\beta}}\bigg|_{\tilde{t},\tilde{s}}\xi_{\beta}^{(n)}
\label{eq5}
\end{equation}
and
\begin{equation}
\xi_{\alpha}^{(n+1)} = \sum_{\nu=1}^{m} \frac{\partial f_{\alpha}}{\partial t_{\nu}}\bigg|_{\tilde{t},\tilde{s}}\xi_{\nu}^{(n)}
+ \sum_{\beta=1}^{m} \frac{\partial f_{\alpha}}{\partial s_{\beta}}\bigg|_{\tilde{t},\tilde{s}}\xi_{\beta}^{(n)} 
\label{eq6}
\end{equation}
This may be combined to write a matrix equation:
\begin{eqnarray}
\begin{pmatrix}
\xi_{\mu}^{(n+1)}\\
\xi_{\alpha}^{(n+1)}
\end{pmatrix}
=
\begin{pmatrix}
\frac{\partial f_{\mu}}{\partial t_{\nu}} & \frac{\partial f_{\mu}}{\partial s_{\beta}} \\
\frac{\partial f_{\alpha}}{\partial t_{\nu}} & \frac{\partial f_{\alpha}}{\partial s_{\beta}}
\end{pmatrix}
\begin{pmatrix}
\xi_{\nu}^{(n)} \\
\xi_{\beta}^{(n)}
\end{pmatrix}
= J 
\begin{pmatrix}
\xi_{\nu}^{(n)} \\
\xi_{\beta}^{(n)}
\end{pmatrix}
\label{eq8}
\end{eqnarray}

Here $J$ is the stability matrix.
Eq. (\ref{eq8}) forms a linearized map. The perturbed eigenvectors can be written as $\xi^{(n)} = e^{n\lambda}\xi^{(0)}$, where $\lambda$ is the generalized
Lyapunov exponent. Henceforth, we will simply call it as Lyapunov exponent. The 
stability matrix eigenvalue equation takes the form 
$J\xi^{(0)} = \sigma\xi^{(0)}$, where $\sigma=e^\lambda$ is the eigenvalue of the stability matrix and $\xi^{(0)}$ is the eigenvector.

From the nonlinear dynamics perspective, the iteration procedure may be viewed as
a discrete time propagation of the molecular ground state under the influence of an input perturbation.
This time-discrete linear state-space model is known to be exponentially stable if all 
the eigenvalues of $J$ have a modulus smaller than one. Note that this analysis is solely based on
linearization, and thus tells us nothing about the marginal cases, in which the neglected terms of the 
order $O({\xi^{(n)}}^2)$ determine the local stability of the system under time-discrete propagation.

It is true that the fixed-point iteration in Eq. (\ref{eq1}) for standard 
CC is rarely used since its convergence is often not so good. Instead, DIIS acceleration
is almost universally used.\cite{pulay1980convergence, scuseria1986accelerating} However, we reiterate that the fixed point iteration procedure
often reveals interesting dynamics associated with the CC iteration time series and 
further reveals interesting mutual inter-dependence and interrelationships among the cluster 
operators. In this 
regard, one may imagine the regularization parameter $\eta$ as an input probe to study the 
extent of non-linearity of the dynamics. As such, it indicates the effect of coupling among them. 
Using thresholded and unthresholded recurrence analysis, 
we will show in sec. \ref{section: Results} that this dynamics of 
the iteration procedure is dictated almost solely by a small subgroup of excitations with large 
amplitudes which involve chemically active orbitals. The
smaller amplitudes of all other excitations are governed solely by the previous subgroup of 
amplitudes. In what follows, we will show in sec. \ref{section: Implications} 
that one may exploit this feature to design a CC theory where the iteration procedure can be
restricted to determine only a few large cluster amplitudes, whereas all other smaller cluster amplitudes
may be determined through a fixed functional dependence on the former subset. Finally, we will 
summarize our findings in sec \ref{section: Conclusion}.

\section{Results}
\label{section: Results}
%\textbf{\textit{Results:}}
The stability of the iterative procedure depends upon the eigenvalues of
the associated stability matrix. If all the eigenvalues of the stability matrix are less than one (i.e., the corresponding Lyapunov exponents are negative), then the procedure converges. Thus $|\sigma|<1$ is the convergent condition for any iterative procedure. A detailed study of the highest Lyapunov exponents of each symmetry for symmetrically stretched water (bond length = 2.6741 Bohr, bond angle = 96.774\degree, cc-pVDZ basis) is reported in Fig. 1. It is shown that the highest Lyapunov exponent (corresponding to $A_1$ symmetry) becomes positive at $\eta = 0.29$.
\begin{figure}[!htbp]
%\centering
\hglue -1.2cm
         \includegraphics[width=\linewidth]{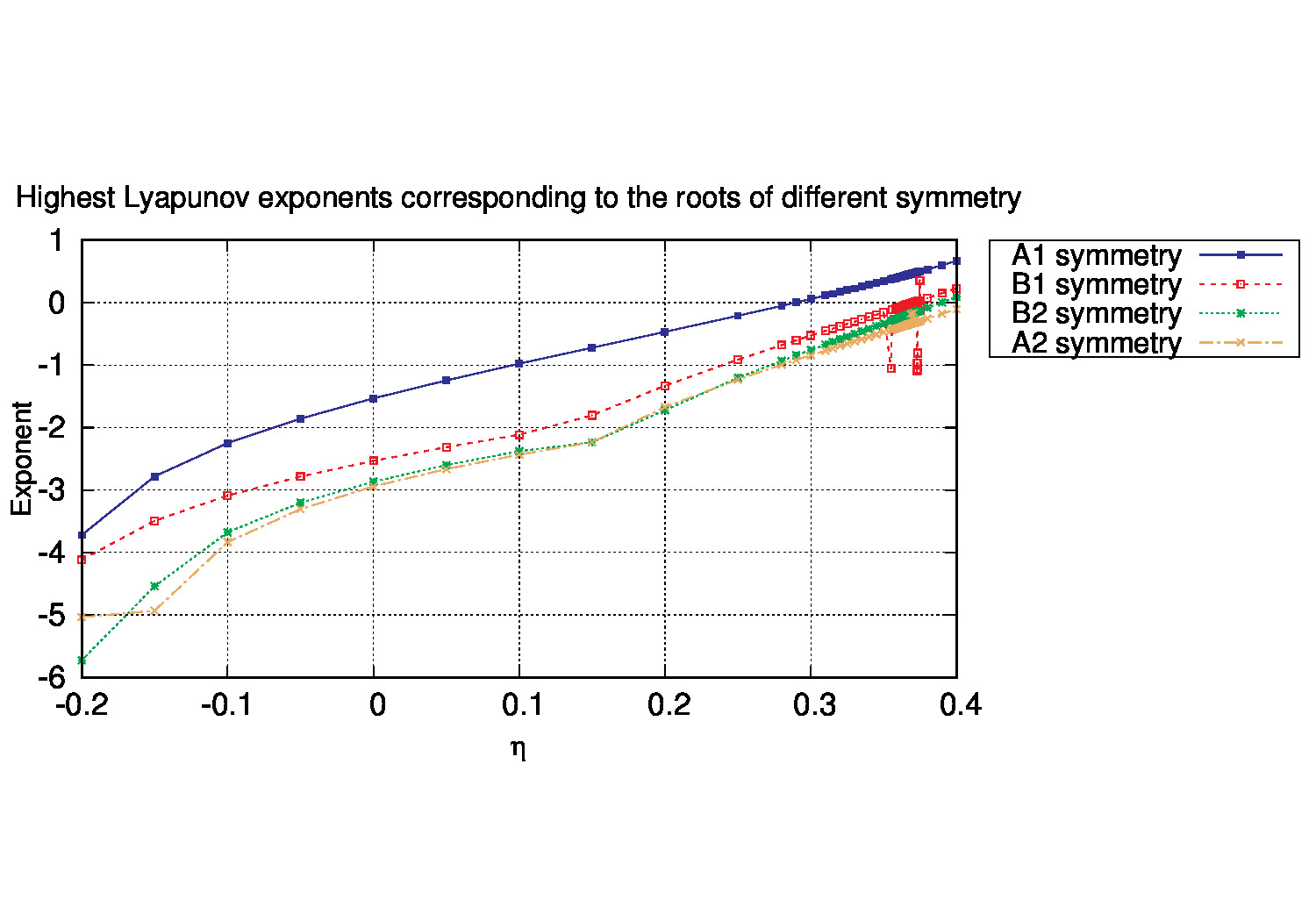}\\
\vspace{-0.8cm}\
\captionsetup[figure]{font=small,skip=1pt}
 \caption{The largest Lyapunov exponents of different roots of different symmetry of $H_2O$ in cc-pVDZ basis for $\eta$ values from -0.2 to 0.4}
 \label{res1}
 \end{figure}

Near the point of equilibrium with small enough $\eta$, the system is Lyapunov stable and the iteration 
converges to the same set of fixed  points\cite{malkin1952theory}. In fact, there is a 
range of $\eta$, for which the system takes fewer number of steps to 
converge, after which it increases sharply.
Here we quantify the effects of larger input disturbances, as done in control theory.

\begin{figure*}
  \includegraphics[width=\textwidth]{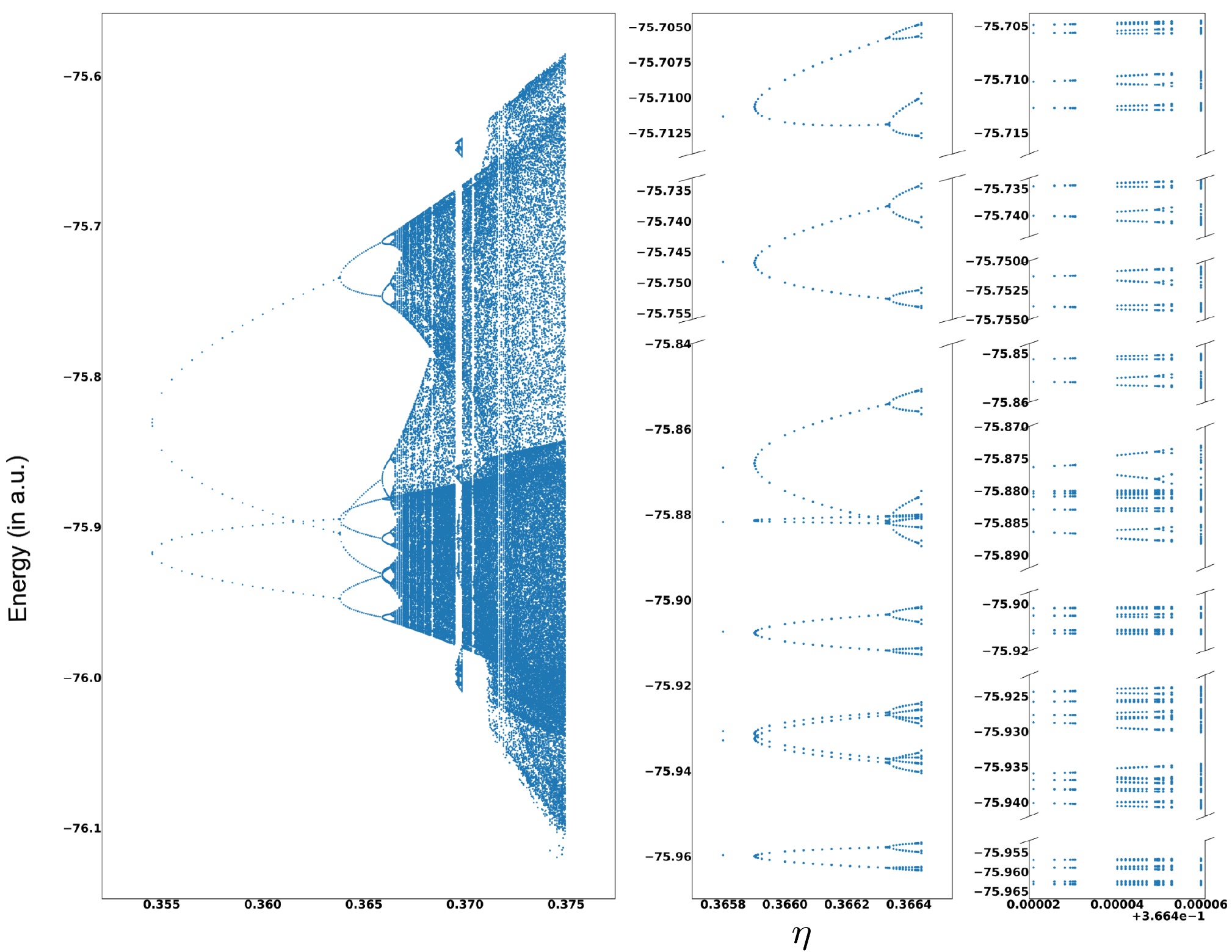}
  \caption{The bifurcation Diagram. Magnified plots in the middle and right columns show clear presence of 16, 32, 64, 128 etc period cycles.}
  \label{fig:Pitchfork}
\end{figure*}

Around $\eta=0.2712$, the perturbation crosses the critical value and one observes the onset of an 
oscillatory divergence in the initial phase of the iteration process, followed by the generation of
period-2 cycle. $\eta$ may be considered as a measure of the non-linearity in the system. 
However, a linear Lyapunov 
stability analysis is unable to predict the cases where the perturbation is large. 
For our coupled time-discrete map, the severe non-linearity results in an early onset of period-2 cycles. With increasing value 
of the parameter, $\eta$, one further observes period-$2^n$ cycles ($n=2, 3,...$), before the iteration
becomes chaotic. Note that for such 1-dimensional dynamics, a full period doubling cascade \textit{must} precede chaos\cite{yorke1985period}. The cluster amplitudes at an arbitrarily chosen $k$-th step recur at $(k+2^n)$-th step for period-$2^n$ cycle and the energy obtained by evaluating the vacuum expectation value $\langle e^{-T} W  e^T\rangle$ oscillates between $2^n$ periods. For a range of $\eta$, it shows period-doubling bifurcation cascade (Fig. \ref{fig:Pitchfork}).

The range of the parameter for successive higher period cycles keeps on shrinking as a characteristic of the period-doubling bifurcation\cite{feigenbaum1978quantitative}. In fact, in the limiting case, i.e., at the onset of chaos, any single parameter map follows the dynamics such that 
\begin{equation}
  \lim_{n \to \infty} \delta \approx 4.6692  , \;\; \; \;  \delta=\frac{\eta_{n+1}-\eta_{n}}{\eta_{n+2}-\eta_{n+1}} 
\end{equation}
Here $\eta_n$ is the onset point of the period-$2^n$ cycle. The limit of $\delta$ is a universal constant, known as Feigenbaum constant. Despite being a multivariate map with tensorial structure, it is indeed possible to generate all the different period cycles, as shown in Fig. \ref{fig:Pitchfork}. 
With a numerical algorithm based on the bisection method, we have precisely determined the 
onset points of different period cycles.
The first eight values of $\eta$ are shown in Table \ref{tab:my_label}, along with the ratio $\delta$. In the limiting case of $n \to \infty$, the $\delta$ computed shows excellent (error\textless 0.004\%) agreement to the universal value of the Feigenbaum constant. Thus, the iteration process under an perturbation, like general n-dimensional 
maps of 1-parameter family\cite{collet1981period}, show the bifurcation doubling route to chaos, with the same geometric 
rate given by the Feigenbaum constant.
The very high value of the first ratio demonstrates an early onset of period-2 cycle, as otherwise predicted by the largest Lyapunov exponent. Thus, it tells us about the severe nonlinearity of the system. One should further note that the convergence to the exact value with higher period cycles is not monotonic, rather we observe an oscillating convergence.
\begin{table}[!b]
    \centering
    \begin{tabular}{|c|c|c|c|}
    \hline
    n & Period ( $=2^n$)  & $\eta$ & $\delta$ (\% error in $\delta$) \\ \hline
    1 & 2                 & 0.2711953  & --     \\ \hline
    2 & 4                 & 0.3545298  & --     \\\hline
    3 & 8                 & 0.363806   & 8.983(92.4\%)  \\\hline
    4 & 16                & 0.3659015  & 4.428(5.2\%)  \\\hline
    5 & 32                & 0.3663366  & 4.8147(3.1\%) \\    \hline
    6 & 64                & 0.36642994 & 4.66145(0.165\%)\\\hline
    7 & 128               & 0.366449917 & 4.67237(0.068\%) \\\hline
    8 & 256               & 0.3664541953  & 4.66937(0.00364\%) \\\hline
    \end{tabular}
    \caption{Onset points of different period cycles and Feigenbaum Constant}
    \label{tab:my_label}
\end{table}

Such clear separation of different periodic cycles, which is characteristic of symmetric few variable systems, indicates the presence of a set of few variables which govern the dynamics. In cases where the linear stability is lost, it is indeed possible to eliminate most of the degrees of freedom from the non-linearly interacting subsystems, so that the macroscopic behavior of the system is governed by a few degrees of freedom only \cite{Haken1983, haken1982slaving, Haken_1989}. Along this line, we presume that the dynamics of our system is dictated by \textit{only a few} large cluster amplitudes, which are the order parameters (unstable modes) of the system that determine its macroscopic pattern. 

We have further studied the dynamics with recurrence analysis\cite{poincare1890probleme, eckmann1987recurrenceplot}. Here each iteration was embedded as one time step\cite{zbilut1992embeddings}. The state $x_i$ was taken as a vector comprising of all the amplitudes at the $i$-th time step;$x_i=(t_{1,i} \oplus t_{2,i})^T$, where $t_{1,i}$ and $t_{2,i}$ denote
the one and two body cluster amplitudes at the $i$-th time step. Distance matrix (DM, also known as the unthreshold recurrence matrix), which is a useful tool to study the phase space trajectory, is defined as 
\begin{equation}
    DM_{i,j} = ||\Vec{x_i}-\Vec{x_j}||,\;\;\;\;i,j = 1,...,N
\end{equation}{}
where $N$ is the Length of time series, and $||\cdot||$ is a norm. Here the simulation was run for 4000 steps and plotted for the last $N=64$ iterations.

\begin{figure}
  \includegraphics[width=\linewidth]{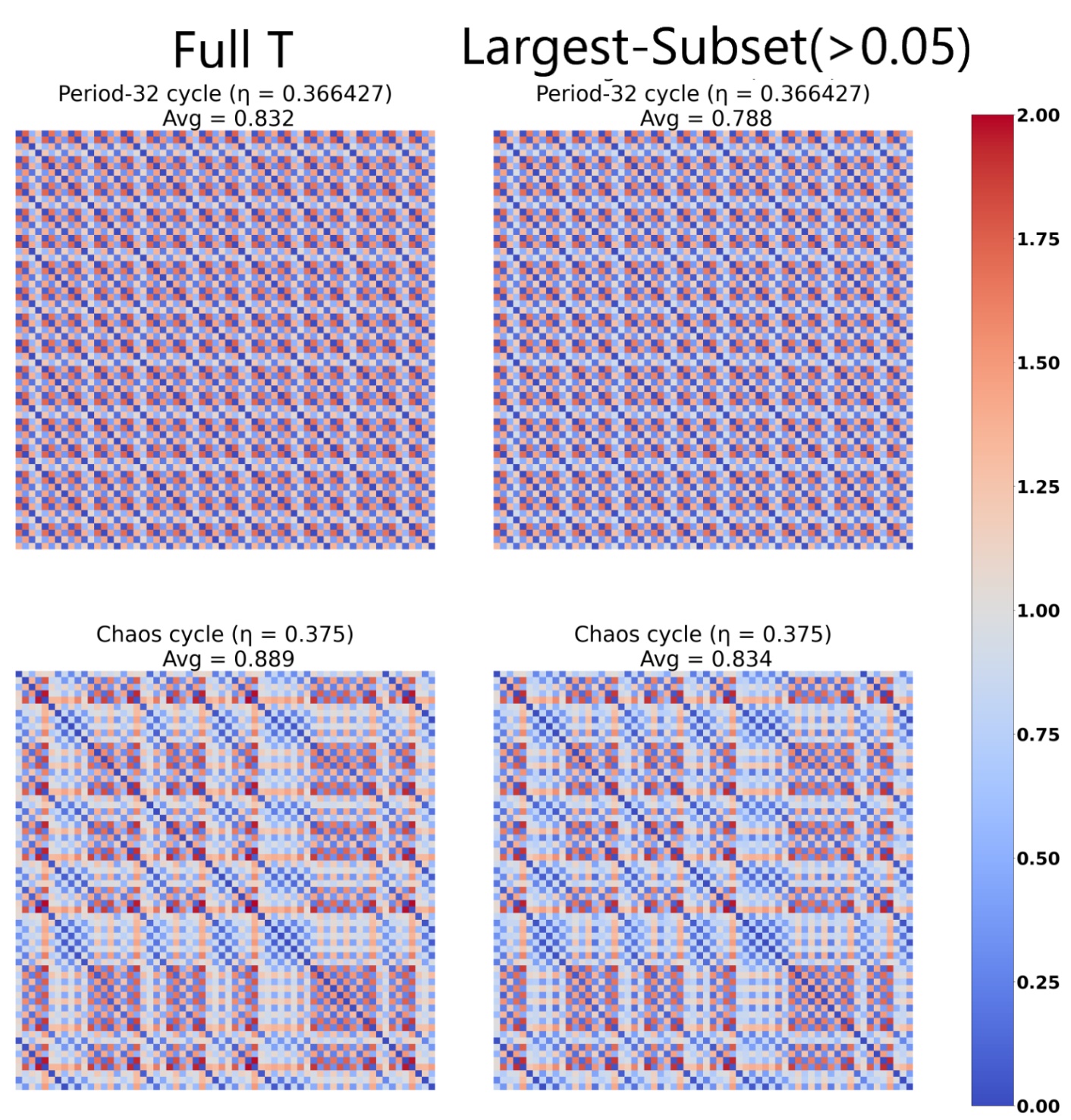}
  \caption{Distance Matrix for a few $\eta$ values with Full-T amplitudes (left column) and Largest-subset of amplitudes, $\{t^L\}$, (right column). Both the horizontal and the vertical axes represent discrete time steps.}% The average variation calculated as $ \frac{1}{N^2} \sum_{i,j}^N DM_{i,j};  \;\;N=64.$ }
  \label{fig:RM}
\end{figure}

\begin{figure}
    \centering
    \includegraphics[width=\linewidth]{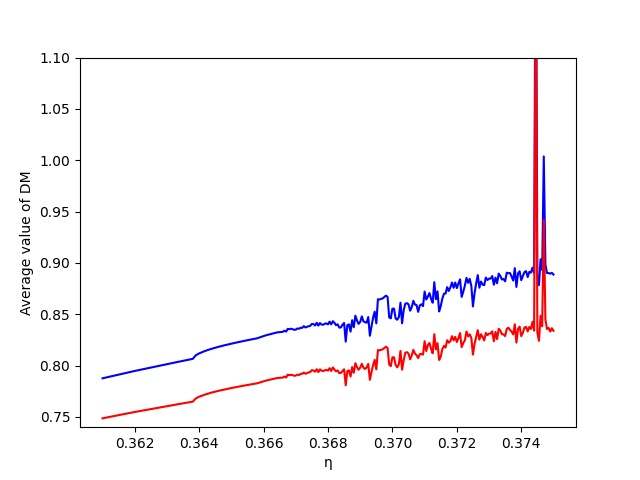}
    \caption{Average variation, calculated as $ \frac{1}{N^2} \sum_{i,j}^N DM_{i,j};  \;\;N=64$, of the cluster amplitudes for the largest subset (red) show qualitatively identical pattern to that of the full-T (blue).}
    \label{fig:averageDM}
\end{figure}

We have plotted DM taking all the (non-zero) cluster amplitudes, here on referred as 
Full-T (4938 non-zero variables), and is shown for some representative $\eta$ values 
(Left panel of Fig. \ref{fig:RM}). To support our hypothesis stated before, we have
divided the cluster amplitudes into two subsets: the largest subset, denoted by
$\{t^L\}$ and the smaller subset, $\{t^S\}$.
We have identified a set of few large cluster operators (total eleven) which have amplitudes 
\textgreater 0.05 throughout the entire range of $\eta$. These operators involve at 
least a pair of chemically active orbitals, and belong to the largest-subset $\{t^L\}$. Rest of the amplitudes are elements of the smaller subset $\{t^S\}$.
It is found that the DM for different $\eta$, constructed with 
the Largest-subset of amplitudes, $\{t^L\}$, replicates the phase space trajectory 
to that obtained 
by Full-T with excellent quantitative and qualitative accuracy. This is shown in the 
right panel of Fig \ref{fig:RM} at the same $\eta$ values. The average variation for the 
largest-subset, $\{t^L\}$,  matches to that constructed with Full-T with 
\textgreater93\% accuracy through the entire range of $\eta$ as shown in Fig. \ref{fig:averageDM}. 
Furthermore, the variation of the average value of the DM constructed by the full set of t-amplitudes
is qualitatively replicated by the largest subset.
Thus the variation of the smaller amplitudes is almost entirely suppressed by the largest-subset, making them asymptotically negligible in the dynamics. The macroscopic pattern is solely determined by the dynamics of the large amplitudes which behave as the order parameters of the system. Their variation is independent of the microscopic sub-dynamics of the smaller amplitudes. In other words, the largest subset enslaves the smaller amplitudes. Such domination of the large amplitudes is amplified by the non-linear terms of the CC expansion, whereas the small amplitudes effectively contribute at the linear level to provide the feedback coupling.

One may wonder whether the dynamics would repeat if the small amplitudes are set to zero. 
We reiterate that it is absolutely crucial that the small amplitudes provide the feedback coupling 
in determining the larger amplitudes, while the former set of amplitudes are determined by the
latter solely. In Synergetics, this is known as the circular causality. The internal dynamics of the
smaller amplitudes can be considered to be frozen.

\begin{figure}
%\centering
    \includegraphics[width=\linewidth]{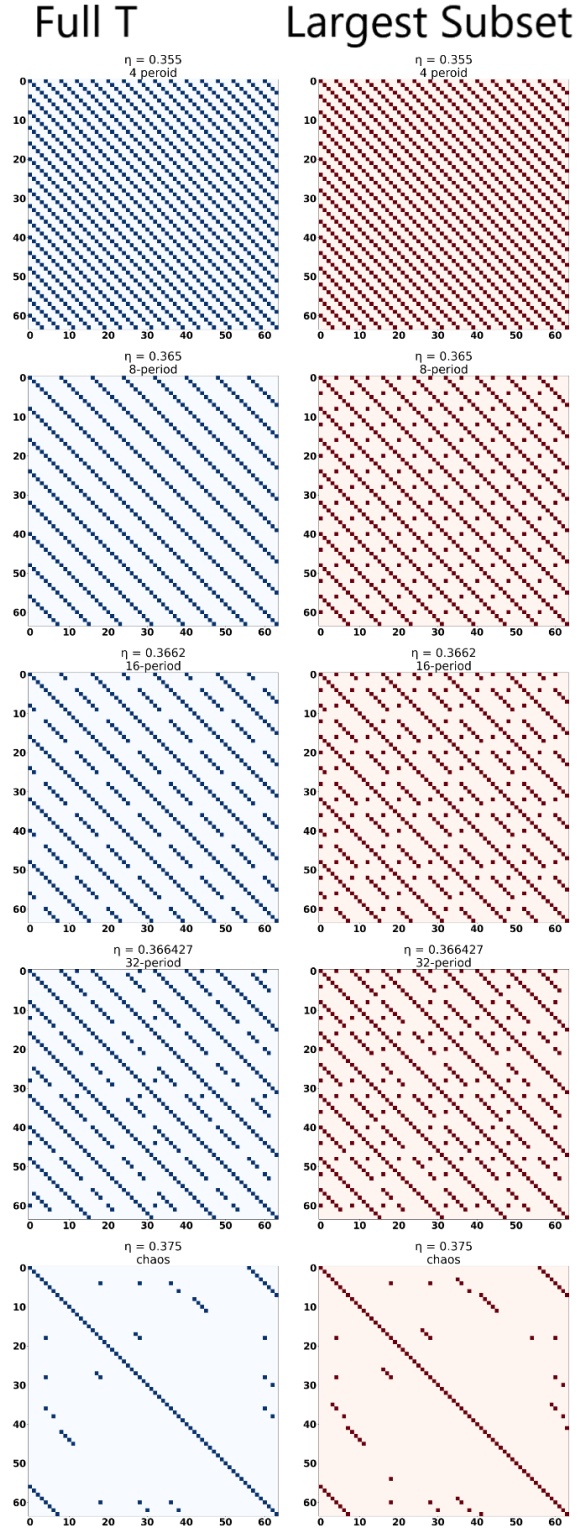}
    \caption{Recurrence Plots for a few $\eta$ values, and threshold $\epsilon = 0.05$ with Full-T amplitudes (left column) and Largest-subset of T-amplitudes (right column). Both the horizontal and the vertical axes represent discrete time steps. }
    \label{fig:RP_main}
\end{figure}
\begin{figure*}
  \includegraphics[width=\textwidth]{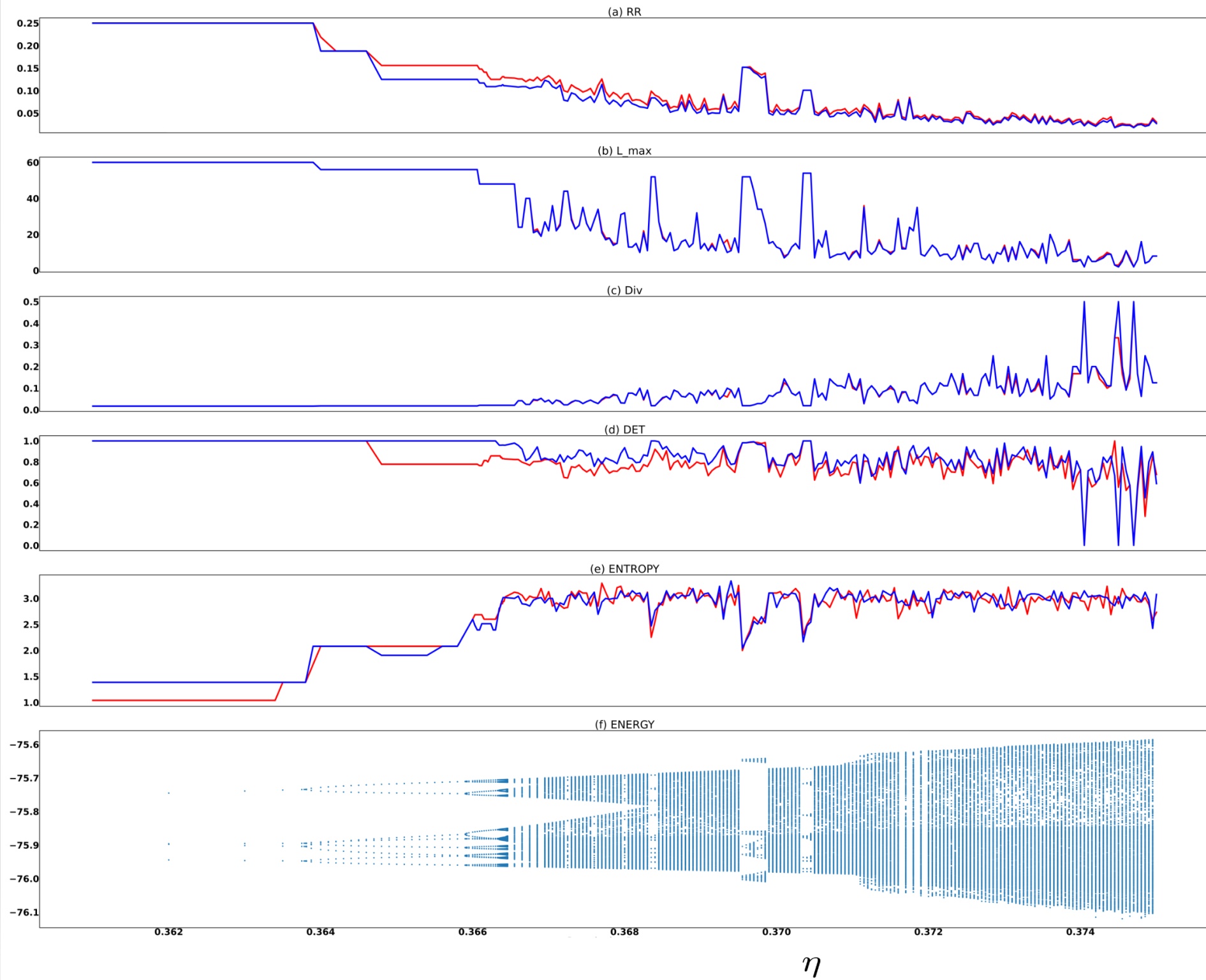}
  \caption{Variation of (a) RR, (b) $L_{max}$, (c) DIV, (d) DET, and (e) Entropy as a function of the regularization parameter $\eta$ computed via RQA. Blue line denotes the quantities with Full-T, and red line denotes that of Largest-subset. The bifurcation diagram (f) is also presented along the same horizontal scale to identify the period-period and chaos-period transitions locations.}
  \label{fig:RQA}
\end{figure*}
\subsection{Recurrence Quantification Analysis (RQA)}
\label{section: RQA}
Recurrence plot (RP) is heuristic approach\cite{marwan2007recurrence,RecurrencePlot} 
to quantify the epochs of a particular state to recur 
in a time-series and is based on its phase space trajectory. A recurrence 
matrix is defined as 
\begin{equation}
RP_{i,j} = \Theta(\epsilon_i - DM_{i,j})  
\end{equation}
where $\epsilon_i$ is a suitable threshold distance, and $\Theta(\cdot)$ is the 
Heaviside function. Thus if the $DM_{i,j}$ is less than the threshold, the 
corresponding $RP_{i,j}= 1$ and denoted by a black dot, otherwise $RP_{i,j  }= 0$ 
(white dot). In RQA, one quantifies the density of recurrence points as well as 
the histograms of the lengths $l$ of the diagonal based on a suitably chosen 
recurrence threshold $\epsilon$. The recurrence threshold is the most 
significant quantity in RQA. It should be chosen small enough to distinguish 
the closely spaced trajectories but not small enough to miss out on the rich 
dynamics associated with the time series. We have chosen $\epsilon_i=0.05$ 
for all further analyses. 
RPs at a few selected values of $\eta$ with $\epsilon = 0.05$ is given 
in Fig (\ref{fig:RP_main}), where the left blue panels represent those constructed
with the full set of t-amplitudes, while the right red panels are constructed
with the largest subset. In both the cases, for a given $\eta$, the RPs are 
characterized by similar continuous diagonal lines, although the RPs constructed 
with the largest subset (right red panels) sometimes may have denser 
population of isolated black dots. 

Recurrence Analysis gives us quantitative measures of different quantities associated
with the dynamics. 
Recurrence rate (RR), which is a measure of the density of recurrence points 
and signifies the probability of occurrence of a specific state, is defined as 
\begin{equation}
    RR=\frac{1}{N^2}\sum_{i,j=1}^{N}RP_{i,j}
\end{equation}
Thus, higher RR corresponds to a more repetitive state space trajectory.  
Fig. \ref{fig:RQA}(a) shows the RR of our system and displays a gradual decrease 
from lower period cycles to higher periodic cycles and eventually to chaos. 
Note that the RR obtained from the time evolution of the largest subset, $\{t^L\}$, 
(red plot) follows quantitatively closely to that obtained from Full-T (blue plot). 
One may note that the choice of the threshold is sensitive enough to capture 
chaos-period transition in the islands of stability around $\eta=0.369$ 
and $\eta=0.372$, which are characterized by sudden upward spikes in the RR plot.    

Deterministic periodic systems are often characterized by repeated long and 
continuous diagonal lines in their RPs, signifying repeated similar state evolution. 
The RPs corresponding to fewer period cycles have larger number of continuous 
diagonal lines parallel to the Line of Identity (LOI) in a given length of the time 
series. Contrarily, subsequent independent values often appear as isolated single 
points. Thus the fraction of recurrence points appearing as diagonal points parallel to 
the LOI is considered to be a measure of determinism of the system: 
    \begin{equation}
    DET=\frac{\sum_{l=l_{\text{min}}}^{N}lP(l)}{\sum_{l=1}^{N}lP(l)}
\end{equation}, where $P(l)$ is the histogram of diagonal lines of length $l$. 
$DET$ is a measure of predictability of the system. As a necessary 
(but not sufficient) condition, the periodic systems are characterized by 
high value of DET and this has successfully been predicted quantitatively 
by the RQA, both with Full-T (blue) and Largest-subset (red) of T amplitudes 
(Fig. \ref{fig:RQA}d).
    
High value of Maximum Diagonal Length ($L_\text{max}$), defined as $L_{\text{max}} = \text{max}({l_i; i=1,2,...,N_l})$ is often characteristic of regular, correlated and periodic systems. In the RQA, one may roughly interpret its inverse, known as Divergence ($DIV$), defined as $ DIV=L_{\text{max}}^{-1}$, as an estimator of Lyapunov Exponent\cite{trulla1996recurrence}. Excluding the LOI and an appropriate Theiler window\cite{theiler1990estimating} around it, the other recurrence points from the subsequent phase space vectors, lead to continuous diagonal lines in the RP. Thus, the lower period cycles, with frequent recurrence of the same states, have higher $L_{max}$ value and lower DIV. This has been quantitatively verified and shown that these two measures have identical behavior when full-T amplitudes and Largest-subset are used (Figs. \ref{fig:RQA}(b) and \ref{fig:RQA}(c)).

One of the important quantities which emerges from RQA is the entropy associated with the dynamics. However, studies have shown that the using Shannon Entropy obtained from RP using its diagonal length histograms, given by $ENTR = -\sum_{l=l_{\text{min}}}^Np(l)ln(p(l))$, where $p(l)$ is the probability distribution of the diagonal length, gives a counter-intuitive trend of entropy for period-chaos systems \cite{letellier2006estimating}. It is observed that such a description often show decreasing value of entropy with increasing chaos and also it anti-correlates with the maximal Lyapunov exponent. A number of methods have been suggested for calculation of entropy\cite{ letellier2006estimating,corso2018quantifying,Weighted_Entropy}. Following Eroglu \textit{et. al.}\cite{Weighted_Entropy}, we have calculated it from Weighted Distance Matrix(WDM),  defined by
\begin{equation}
  {W}_{i,j}=e^{-||\Vec{x_i}-\Vec{x_j}||}
\end{equation}
For entropy calculation, we define strength ($s_i$) as \newline $s_i=\sum_{j=1}^N W_{i,j}$. 
The strength is used to calculated Shannon Entropy associated with the WDM through the distribution of P(s). 
\begin{equation}
    ENTR=-\sum_{\{s\}}p(s)ln \;p(s)
\end{equation}
where $p(s) = P(s)/S$ is the relative frequency distribution of WDM and $S=\sum_i^N s_i$. The variation of $ENTR$ with respect to $\eta$ is shown in Fig. \ref{fig:RQA}(e). Clearly this correlates with the Lyapunov exponent and shows all the jumps and dips of period-period and chaos-period transitions across different range of $\eta$. The $ENTR$ predicted with the Full-T amplitudes and that with the largest subset follow closely throughout, thus $ENTR$ is shown to be governed solely by the active excitations.

It is thus evident that for all the measures of the RQA, the few large amplitude 
excitations containing the active orbitals qualitatively and quantitatively replicate those
shown by the entire set of cluster operators. In other words, the internal 
dynamics of thousands of the smaller amplitudes does not effectively contribute to the 
iteration dynamics of the entire system, and hence it can be considered to be frozen.
All such RQA measures are found to be largely independent of the time-series embedding dimension.
In the next subsection, we will show that such interrelationship among different 
sets of cluster amplitudes may be exploited to construct an effective CC theory.

\subsection{Implications to Quantum Chemistry}
\label{section: Implications}
While the chaotic dynamics encountered in quantum chemistry methods is
interesting in itself, it has immense implications to gain insight into methods of quantum chemistry. 
Note that the principles of Synergetics allow us to map each of the small amplitudes as unique 
functions of the larger amplitudes. In other words,
\begin{equation}
t^S_k=F_k(\{t^L\})
\end{equation}
where $t^S_k$ is the $k$-th element of the smaller subset $\{t^S\}$ and $F_k$ is the functions
of the largest subset which determines the $k$-th element of the smaller subset. 
In the absence of any internal 
dynamics of the smaller subset, the above equation implies that each element in the smaller subset is a
function of the largest subset only. The functional dependence of the smaller subset on the largest subset
is fixed through the entire iteration process. That implies that one may devise an effective CC theory
where the iteration may be restricted to only a few degrees of freedom. One may 
iteratively update the residues of \textit{only the few excitations belonging to the largest subset} 
to determine the corresponding cluster amplitudes, whereas the smaller amplitudes couple in their
equations to provide the feedback coupling in the iteration process. One may further accelerate 
the iteration process via usual numerical techniques, like DIIS. Hundreds of smaller amplitudes, 
on the other hand, may be computed from their predetermined functional map without going through 
the iterative procedure.
This then reduces the computational scaling drastically. Furthermore, the dimension of the 
largest subset $\{t^L\}$ is independent of the size of the basis set. This constitutes the basis of an 
effective CC theory with reduced dimensionality, which is currently under development and warrants 
a separate publication elsewhere.

\section{Conclusion}
\label{section: Conclusion}
%\textbf{\textit{Conclusion}}: 
In summary, we have shown that the discrete-time propagation 
associated with the iterative procedure of a double similarity transformed CC theory 
shows the dynamical features of a logistic map. The dynamics shows the usual 
period doubling bifurcation route to chaos, with the universal geometric rate given by the 
Feigenbaum constant. Further to that, recurrence analysis was performed with the 
state vectors comprising all the cluster amplitudes and that with the largest subset thereof. The RQA 
shows identical phase space trajectory for these two cases. This reinforces our hypothesis that only a few of the excitations with large amplitudes govern the iteration dynamics, whereas hundreds of 
cluster operators with smaller amplitudes are enslaved. 

Given the fact that the internal dynamics of the smaller
amplitudes are frozen, we propose a novel version of CC theory where one may restrict the iteration 
procedure only to determine the amplitudes of the largest subset, whereas the smaller amplitudes
may be determined from their functional dependence on the largest subset. Thus the principles of
Synergetics is likely to pave the way to develop a few variable effective CC theory.
\begin{acknowledgments}
% Put your acknowledgments here.
RM acknowledges IIT Bombay Seed grant, and Science and Engineering Research Board, Government of India, for financial support. Discussions with Professors Debashis Mukherjee and Sandip Kar are acknowledged.
\end{acknowledgments}
\section*{Data Availablity}
The data that support the findings of this study are available from the corresponding author upon reasonable request.

%\nocite{*}
%\bibliography{literature}% Produces the bibliography via BibTeX.

\end{document}